\documentclass[10pt,conference]{IEEEtran}
\IEEEoverridecommandlockouts
\usepackage{cite}
\usepackage{amsmath,amssymb,amsfonts}
\usepackage{algorithm}
\usepackage{algpseudocode}
\usepackage{mdframed}
\usepackage{graphicx}
\usepackage{textcomp}
\usepackage{xcolor}
\usepackage{physics}
\usepackage{tikz}
\usetikzlibrary{quantikz2}
\usepackage{float}
\usepackage{hyperref}
\usepackage{xcolor}
\usepackage{balance}

\def\BibTeX{{\rm B\kern-.05em{\sc i\kern-.025em b}\kern-.08em
 T\kern-.1667em\lower.7ex\hbox{E}\kern-.125emX}}

\algdef{SE}[FOR]{ForAll}{EndForAll}[1]{\algorithmicforall\ (#1)}{\algorithmicend\ \algorithmicfor}%
\algdef{SE}[IF]{If}{EndIf}[1]{\algorithmicif\ #1}{\algorithmicend\ \algorithmicif}%
\algdef{C}[IF]{IF}{ElsIf}[1]{\algorithmicelse\ \algorithmicif\ #1}%

\begin{document}

\bstctlcite{IEEEexample:BSTcontrol}

\title{\huge Branch-Resolved Characterization of Feed-Forward Error in Dynamic Teleportation via Classical Choi Shadows
}

\author{\IEEEauthorblockN{Mason Edwards and Prabhat Mishra}
\IEEEauthorblockA{
\textit{University of Florida, Gainesville, Florida, USA} \\
}
}

\maketitle

\begin{abstract}
Mid-circuit measurement and classical feed-forward are essential primitives for dynamic-circuit teleportation on superconducting quantum processors. However, the error associated with measurement-conditioned corrective operations remains poorly understood when evaluated with respect to individual measurement branches. In this paper, we present a framework for characterizing feed-forward error in dynamic circuit teleportation without losing valuable information related to its behavior across separate branches. We analyze three approaches to applying measurement-conditioned corrections: (i) physical application, (ii) post-processing adjustments, and (iii) a mitigated physical application which utilizes Bit-Flip Averaging (BFA)-based Probabilistic Readout Error Mitigation (PROM). We experimentally reconstruct branch Choi operators via an entangled reference qubit, and validate our physical-application and post-processing Choi-shadow estimators against full tomography of the branch Choi operators. We perform experiments on two physical qubit layouts which differ greatly in mid-circuit measurement readout error, and observe a reversal in the relative order in branch qualities obtained from the post-processing and PROM mitigation strategies. In one physical layout with higher measurement readout error, the operational feed-forward penalty is relatively modest (approximately 0.02-0.03) and PROM produces higher branch qualities than post-processing for every branch. In a separate layout with lower readout error, the operational feed-forward penalty increases to roughly 0.07, and post-processing exceeds PROM for all branch qualities. Our characterization framework can reveal branch-specific error structure and mitigation behavior that state-of-the-art outcome-averaged analyses fail to expose.
\end{abstract}


\section{Introduction}
\label{sec:intro}

The emergence of dynamic circuit features in Noisy Intermediate-Scale Quantum (NISQ) computers has dramatically changed the types of computational processes that can be performed~\cite{volya2023quantum,  
volya2024quantum, 
volya2021quantum,
sanjaya2026controlled, 
sanjaya2024variational}.
For example, recent work has demonstrated that dynamic circuits with mid-circuit measurement and feed-forward can reduce the amount of resources required to perform common operations including the quantum Fourier transform~\cite{Baumer_Tripathi_Seif_Lidar_Wang_2024}, and enable shallow-depth implementations~\cite{Cao_Eisert_2026}. Before the introduction of mid-circuit measurement and classical feed-forward, conventional quantum circuits only offered terminal measurements, which do not provide support for altering future operations based on their outcome. Many endeavors in quantum computing, including error correction, state preparation, qubit reset, and qubit reuse, depend on the ability to conditionally execute certain quantum operations based on the classical readouts~\cite{Koh_Koh_Thompson_2026, DeCross_Chertkov_Kohagen_Foss-Feig_2023, volya2024state, volya2023feedback}. However, these efforts face two challenges. Since the measurement process itself can lead to erroneous readout~\cite{utt2024quantum}, it can prevent the correct feed-forward operation from being utilized, obscuring the intended post-measurement behavior of the circuit. Moreover, mid-circuit measurements produce both classical and quantum outputs~\cite{Davies_Lewis_1970, Stricker_Vodola_Erhard_Postler_Meth_Ringbauer_Schindler_Blatt_Muller_Monz_2022, Rudinger_Ribeill_Govia_Ware_Nielsen_Young_Ohki_Blume-Kohout_Proctor_2022}, and effectively branch the system's evolution into separate outcome-defined branches. This added layer of complexity makes it difficult to characterize error related to measurement-conditioned feed-forward operations.

Quantum teleportation provides a convenient platform to study feed-forward error. In this process, the quantum information of an unknown quantum state is transferred from a sender to a receiver. These two share an entangled resource state, and the sender performs a Bell measurement on the unknown state and their part of the resource. The outcome of this Bell measurement is then classically communicated to the receiver, who performs an outcome-dependent corrective quantum operation to obtain the teleported state~\cite{Bennett_Brassard_Crepeau_Jozsa_Peres_Wootters_1993}. The Bell measurement dynamically evolves the composite quantum system into distinct outcome-defined branches~\cite{Rudinger_Ribeill_Govia_Ware_Nielsen_Young_Ohki_Blume-Kohout_Proctor_2022, Stricker_Vodola_Erhard_Postler_Meth_Ringbauer_Schindler_Blatt_Muller_Monz_2022}, with a separate outcome-dependent corrective operation associated with each branch. Hence, the measurement-induced branches of the teleportation process offer an especially convenient environment for studying quantum dynamics and errors defined by measurement outcomes. There exists a large body of work applying such a paradigm to various areas of quantum information processing, such as quantum error correction, qubit routing and compilation, and quantum networking~\cite{quantinuum_logical_qubit_teleportation, Baur_Fedorov_Steffen_Filipp_Da_Silva_Wallraff_2012, Kang_Kam_Mooney_Hollenberg_2025, Devulapalli_Schoute_Bapat_Childs_Gorshkov_2024, Rall_Tame_2024, Steffen_Salathe_Oppliger_Kurpiers_Baur_Lang_Eichler_Puebla-Hellmann_Fedorov_Wallraff_2013}. In this paper, we use teleportation as a resource-controlled benchmark to isolate and measure operational feed-forward error with regard to the branching induced by mid-circuit measurements. 


Dynamic teleportation has gained attention in NISQ computing~\cite{Steffen_Salathe_Oppliger_Kurpiers_Baur_Lang_Eichler_Puebla-Hellmann_Fedorov_Wallraff_2013}. Recent studies have demonstrated a large performance gap between performing corrections actively through dynamic circuits compared to passively in post-selection via measurement categorization~\cite{Kang_Kam_Mooney_Hollenberg_2025}. Mid-circuit measurements can be viewed as quantum instruments~\cite{Rudinger_Ribeill_Govia_Ware_Nielsen_Young_Ohki_Blume-Kohout_Proctor_2022} that produce both classical and quantum output, which can be utilized to determine the subsequent conditional operations~\cite{Stricker_Vodola_Erhard_Postler_Meth_Ringbauer_Schindler_Blatt_Muller_Monz_2022}. Related work has developed tomography methods specific to measurements, including conditioned process tomography of measurement back-action~\cite{Blumoff_Chou_Shen_Reagor_Axline_Brierley_Silveri_Wang_Vlastakis_Nigg_et_al._2016}, quantum instrument tomography~\cite{Rudinger_Ribeill_Govia_Ware_Nielsen_Young_Ohki_Blume-Kohout_Proctor_2022, Stricker_Vodola_Erhard_Postler_Meth_Ringbauer_Schindler_Blatt_Muller_Monz_2022}, and quantum non-demolition measurement tomography~\cite{Pereira_Garcia-Ripoll_Ramos_2022, Pereira_García-Ripoll_Ramos_2023}. The viability of classical shadows for efficiently estimating selected properties of dynamical quantum processes without full process tomography has been demonstrated~\cite{Levy_Luo_Clark_2024}. Recent work has developed an in situ randomized benchmarking method for measuring mid-circuit measurement-induced error rates in many-qubit circuits~\cite{Hothem_Hines_Baldwin_Gresh_Blume-Kohout_Proctor_2025}. Mitigation of readout error via bit-flip averaging has been shown~\cite{Smith_Khosla_Self_Kim_2021}, and PROM has shown the ability to reduce the negative effects of readout error in adaptive circuits with feed-forward~\cite{Koh_Koh_Thompson_2026}. 

Existing work on teleportation benchmarking and characterization primarily uses outcome-aggregated estimates such as final-state fidelity, average process fidelity, and measures of entanglement preservation~\cite{Kang_Kam_Mooney_Hollenberg_2025, Baur_Fedorov_Steffen_Filipp_Da_Silva_Wallraff_2012}, which disregard information related to the dynamics specific to individual measurement branches. Prior work has reconstructed process matrices conditioned on individual Bell-measurement outcomes~\cite{Baur_Fedorov_Steffen_Filipp_Da_Silva_Wallraff_2012, Steffen_Salathe_Oppliger_Kurpiers_Baur_Lang_Eichler_Puebla-Hellmann_Fedorov_Wallraff_2013}. However, the reported normalized outcome-conditioned process matrices themselves do not encode empirical branch probabilities. Detailed tomographic methods have been developed for comprehensive characterization of mid-circuit measurements~\cite{Rudinger_Ribeill_Govia_Ware_Nielsen_Young_Ohki_Blume-Kohout_Proctor_2022, Stricker_Vodola_Erhard_Postler_Meth_Ringbauer_Schindler_Blatt_Muller_Monz_2022, Pereira_García-Ripoll_Ramos_2023}, yet scalable in situ benchmarking remains primarily focused on aggregate mid-circuit measurement-induced error rates in many-qubit circuits~\cite{Hothem_Hines_Baldwin_Gresh_Blume-Kohout_Proctor_2025}.


To the best of our knowledge, the existing literature lacks a branch-resolved, process-level method for characterizing feed-forward error in dynamic teleportation. Such a protocol is necessary to illuminate branching-related discrepancies and provide insight into the extent to which these can be improved by mitigation protocols. In this paper, we address this gap by making the following important contributions:
\begin{enumerate}
    \item We propose a branch-resolved method for characterizing feed-forward error in dynamic quantum teleportation.
    \item We design a controlled pair of state-preparation quantum circuits that prepare both maximally entangled and non-maximally entangled 4-qubit $W_4$ states, yet compile to identical circuit depth and gate count while inducing different ideal corrected teleportation channels.
    \item We demonstrate our framework on IBM Fez, and use two physical qubit layouts to study differing measurement readout conditions. We compare three correction strategies: physical application, post-processing adjustments, and PROM-mitigated physical application. 
    \item We experimentally construct branch Choi operators with classical Choi shadows, validate the physical-correction and post-processing estimators against full tomography of the branch Choi operators, and show that modifying the physical layout reverses the relative branch-quality performance between PROM and post-processing.
\end{enumerate}


\begin{figure*}[htbp]
  \centering
  \includegraphics[width=1\linewidth]{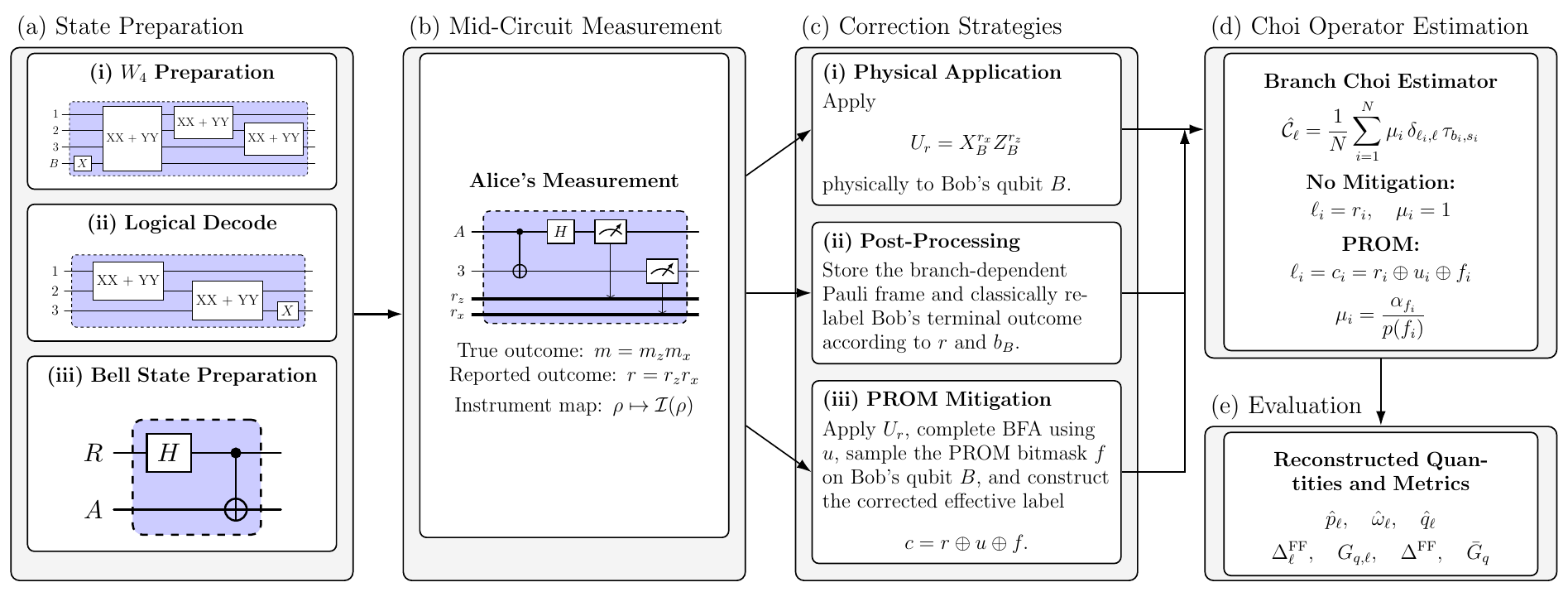}
  \vspace{-0.3in}
  \caption{Branch-Resolved Characterization Framework. $\textbf{(a)}$ Sequence of initial state preparations performed preceding the mid-circuit measurement. First, either entangled 4-qubit $W_4$ resource is prepared on qubits $1, 2, 3, B$ via the quantum circuit shown in Fig.~\ref{fig:resource_prep_circuit}, followed by a decoding of Alice's resource qubits into their logical basis (Eq.~\ref{eq:logical_qubits}) via the quantum circuit shown in Fig.~\ref{fig:logical_decode}. Then, a Bell state is prepared on reference qubit $R$ and input qubit $A$ (Eq.~\ref{eq:bell_state_ra}). $\textbf{(b)}$ Alice's branch-inducing mid-circuit measurement on qubits $A$ and $3$. A preceding bell-basis change is performed, and the measurement outcome readout from qubit $A$ is stored in classical register $r_z$, while the readout from qubit $3$ is stored in classical register $r_x$. Readout assignment error obscures the ability to determine the true branch $m = m_z m_x$ resulting from the measurement. $\textbf{(c)}$ The three methods used to apply the outcome-dependent corrections described in Eq.~\ref{eq:teleportation_corrections}: (i) physical application, (ii) post-processing, and (iii) PROM mitigation. Each method is explained in detail in Section~\ref{sec:experimental_methods}. $\textbf{(d)}$ Estimation of branch Choi operators. Snapshot operators are constructed following terminal measurements in each shot, which are used in our shadow estimator (Eq.~\ref{eq:non_mitigated_estimator}) to estimate the observed branch Choi operators for each measurement-induced branch. For the PROM method, a corrected branch label is used, which targets the ideal branches via our mitigated Choi estimator in Eq.~\ref{eq:mitigated_estimator}. $\textbf{(e)}$ We reconstruct branch probabilities and branch-quality metrics from these shadow estimators, and use them to estimate feed-forward error and the extent of PROM mitigation as detailed in Section~\ref{sec:framework}.} 
  \label{fig:architectural_diagram}
  \vspace{-0.15in}
\end{figure*}

\section{Background: Dynamic Teleportation}
\label{sec:Background}

For teleportation of a single qubit, it is necessary to describe our composite system of qubits in terms of two subsystems, which by convention we refer to as Alice and Bob. Alice's subsystem consists of the input qubit and her part of an entangled resource state which she shares with Bob, whose subsystem also contains the output qubit Alice wishes to teleport the input qubit's state information to. She does this by performing a Bell measurement on her subsystem, which branches the composite system's evolution into one of four possible subchannels. In this paper, we refer to these as branch maps. Alice's measurement outcome $m$ can be represented by two classical bits via $\Omega = \{00,01,10,11\}$, where $m \in \Omega$. In each measurement-defined branch map $\mathcal{E}_m$, the teleported state can be recovered on Bob's post-measurement output qubit via application of a branch-dependent combination of Pauli operators. Using classical feed-forward, Alice thus sends her observed outcome $m$ to Bob, who can then use his knowledge of $m$ to apply the corresponding sequence of Pauli gates to his qubit and recover the teleported input state~\cite{Bennett_Brassard_Crepeau_Jozsa_Peres_Wootters_1993}, in the ideal case where the shared entanglement resource is maximally entangled across the sender-receiver partition~\cite{Agrawal_Pati_2006}.

There are two viable methods for applying Pauli corrections to the output state: (i) physically applying the branch-dependent combination of Pauli gates directly to Bob's qubit, and (ii) making adjustments in post-processing. In this paper, we use two types of four-qubit $W$ states as entanglement resources in our teleportation protocol, building on prior work using $W$ states in teleportation~\cite{Agrawal_Pati_2006, pradhan2007teleportationsuperdensecodinggenuine}. For both resources, the corrective gates for each branch map are: 
\vspace{-0.08in}
\begin{equation}
    U_m = X^{m_x}Z^{m_z},
\label{eq:teleportation_corrections}
\vspace{-0.08in}
\end{equation}
where $\mathbb{P} = \{\mathbb{I}, X, Y, Z\}$ is the set of Pauli operators, and $m = m_z m_x \in \Omega$ labels the true branch obtained from Alice's mid-circuit Bell measurement. We define the physical correction channel as:
\vspace{-0.08in}
\begin{equation}
    \zeta_m(\rho_{\text{post}}) = U_m \rho_{\text{post}} U_m^\dagger,
    \vspace{-0.08in}
\end{equation}
where $\rho_{\text{post}}$ is the uncorrected output state on Bob's qubit following Alice's mid-circuit measurement. This allows us to define the corrected branch map, $\tilde{\mathcal{E}}_m$, as:
\vspace{-0.09in}
\begin{equation}
    \tilde{\mathcal{E}}_m = \zeta_m \circ \mathcal{E}_m.
    \vspace{-0.09in}
\end{equation}

Different mathematical tools are needed for describing the aggregate corrected channel $\tilde{\mathcal{E}}$ and the branch maps $\mathcal{E}_m$. While the aggregate channel can be modeled as a completely positive, trace-preserving (CPTP) map, the mid-circuit measurement branches the system's dynamics into a finite set of completely positive (CP), trace-non-increasing (TNI) maps~\cite{Davies_Lewis_1970,Stricker_Vodola_Erhard_Postler_Meth_Ringbauer_Schindler_Blatt_Muller_Monz_2022, Rudinger_Ribeill_Govia_Ware_Nielsen_Young_Ohki_Blume-Kohout_Proctor_2022}. Since each branch-specific correction $\zeta_m(\cdot) = U_m \cdot U_m^{\dagger}$ is a unitary channel, $\tilde{\mathcal{E}}_m$ retains the CP and TNI properties of $\mathcal{E}_m$. Hence, in our branch-resolved characterization of the teleportation process, we cannot rely solely on a single CPTP-channel description and must instead use the quantum instrument representation to describe the branch-resolved dynamics~\cite{Stricker_Vodola_Erhard_Postler_Meth_Ringbauer_Schindler_Blatt_Muller_Monz_2022, Rudinger_Ribeill_Govia_Ware_Nielsen_Young_Ohki_Blume-Kohout_Proctor_2022}. In this formulation, each outcome $m$ is associated with a CP-TNI map $\mathcal{E}_m$, and for an initial state $\rho_{\text{pre}}$ on Alice's input qubit, the probability of obtaining branch $m$ is given by:
\vspace{-0.05in}
\begin{equation}
    p(m \mid \rho_{\text{pre}}) = \text{Tr} \left[ \mathcal{E}_m \left( \rho_{\text{pre}} \right) \right].
    \vspace{-0.05in}
\end{equation}

The normalized branch output states for each measurement outcome can then be expressed as:
\vspace{-0.05in}
\begin{equation}
    \rho_m = \frac{\mathcal{E}_m(\rho_{\text{pre}})}{p(m \mid \rho_{\text{pre}})},
    \vspace{-0.05in}
\end{equation}
provided $p(m \mid \rho_{\text{pre}}) > 0$. The mid-circuit measurement is therefore represented by a CPTP instrument map $\mathcal{I}$ from a purely quantum input to a hybrid quantum-classical output~\cite{Davies_Lewis_1970, Stricker_Vodola_Erhard_Postler_Meth_Ringbauer_Schindler_Blatt_Muller_Monz_2022}:
\vspace{-0.05in}
\begin{equation}
    \rho \mapsto \mathcal{I}(\rho) = \sum_{m \in \Omega} \mathcal{E}_m(\rho) \otimes \ket{m}\bra{m},
    \vspace{-0.05in}
\end{equation}
where $\ket{m}\bra{m}$ denotes the classical register state storing Alice's two-bit measurement readout used to apply the appropriate correction. Due to the trace-preserving property of $\mathcal{I}$~\cite{Hashim_Nguyen_Goss_Marinelli_Naik_Chistolini_Hines_Marceaux_Kim_Gokhale}, for a normalized input state $\rho$, it follows that:
\vspace{-0.05in}
\begin{equation}
    \sum_{m \in \Omega} \text{Tr}[\mathcal{E}_m(\rho)] = 1.
    \vspace{-0.05in}
\end{equation}
Thus, we can represent the aggregate teleportation process in terms of the branch maps via the expression:
\vspace{-0.05in}
\begin{equation}
    \tilde{\mathcal{E}} = \sum_{m \in \Omega} \tilde{\mathcal{E}}_m.
    \vspace{-0.05in}
\end{equation}

\section{Branch-Resolved Characterization}
\label{sec:framework}

In our framework, we define a multipartite teleportation protocol consisting of six total qubits: one reference qubit, labeled $R$, which we use to construct branch map Choi operators, and five remaining qubits, labeled $A$, 1, 2, 3, and $B$, which we use for teleportation. Qubit $A$ contains the input state $\ket{\psi}_A$ we wish to teleport, while qubits 1, 2, 3, and $B$ contain the prepared entanglement resource state. Motivated by prior work demonstrating the utility of multipartite entanglement in teleportation~\cite{Karlsson_Bourennane_1998}, and by later work showing that non-maximally entangled multipartite states can improve controlled teleportation performance~\cite{Paulson_Panigrahi_2019}, we use two types of entangled four-qubit $W$ states to account for effects related to entanglement resource state structure in our teleportation protocol: (i) a primary resource that is maximally-entangled across the Alice-Bob partition, and (ii) a secondary non-maximally entangled control resource state that is not. The general form of a 4-qubit $W$ state on qubits $1, 2, 3, B$ is:
\vspace{-0.05in}
\begin{equation}
    \ket{\Psi}_{123B} = a\ket{1000} + b\ket{0100} + c\ket{0010} + d\ket{0001}.
    \vspace{-0.05in}
\label{eq:general_w4}
\end{equation} 
To be physically realizable, any valid $W_4$ state must satisfy the normalization condition $\left| a \right|^2 + \left| b \right|^2 + \left| c \right|^2 + \left| d \right|^2 = 1$. The first state we use is the 4-qubit symmetric $W_4$ state, which is defined by its equal amplitudes~\cite{pradhan2007teleportationsuperdensecodinggenuine} $a = b = c = d = \frac{1}{2}$:
\vspace{-0.11in}
\begin{equation}
    \ket{W_4} = \frac{1}{2} \left( \ket{1000} + \ket{0100} + \ket{0010} + \ket{0001} \right).
\label{eq:explicit_symmetric_w4}
\vspace{-0.05in}
\end{equation}
This state is properly normalized, and can be used experimentally. However, Bob's reduced state is not maximally-mixed as $\rho_B = \frac{3}{4}\ket{0} \bra{0} + \frac{1}{4}\ket{1} \bra{1} \ne \frac{\mathbb{I}}{2}$, so the resource has less than one ebit of entanglement across the Alice and Bob partition and thus does not allow for unit-fidelity teleportation~\cite{Agrawal_Pati_2006}. The ideal teleportation channel resulting from its use thus differs from the identity channel. 

The second $W_4$ resource state is chosen so that Bob's reduced state is maximally-mixed. Following the correspondence between maximally entangled resource states, singlet fraction, and perfect teleportation fidelity~\cite{Horodecki_Horodecki_Horodecki_1999}, we refer to this resource as the perfect $W_4$ state. Separate amplitudes can be chosen such that Bob's reduced state is maximally-mixed~\cite{pradhan2007teleportationsuperdensecodinggenuine}. Decomposing Eq.~\ref{eq:general_w4} into Alice and Bob's subsystems yields:
\vspace{-0.05in}
\[
\ket{\Psi}_{123B} = \left(a\ket{100}_{123} + b\ket{010}_{123} + c\ket{001}_{123} \right) \otimes \ket{0}_B
\vspace{-0.05in}
\]
\begin{equation}
+ \;d\ket{000}_{123} \otimes \ket{1}_B.
\label{eq:w4_decomp0}
\vspace{-0.05in}
\end{equation}
We can express this in a form resembling a Bell state by defining unnormalized composite states $\tilde{\ket{\phi_0}}_{123} = a\ket{100}_{123} + b\ket{010}_{123} + c\ket{001}_{123},$ and $\tilde{\ket{\phi_1}}_{123} = d\ket{000}_{123}$ for the qubits in Alice's subsystem. Eq.~\ref{eq:w4_decomp0} then becomes:
\vspace{-0.05in}
\begin{equation}
\ket{\Psi}_{123B} = \tilde{\ket{\phi_0}}_{123} \otimes \ket{0}_B + \tilde{\ket{\phi_1}}_{123} \otimes \ket{1}_B.
\label{eq:unnorm_4_qubit_state}
\vspace{-0.05in}
\end{equation}
The density matrix for such a state can then be expressed as:
\vspace{-0.05in}
\[
\rho = \ket{\Psi} \bra{\Psi} = \tilde{\ket{\phi_0}} \tilde{\bra{\phi_0}} \otimes \ket{0} \bra{0} + \tilde{\ket{\phi_0}} \tilde{\bra{\phi_1}} \otimes \ket{0} \bra{1}
\vspace{-0.05in}
\]
\begin{equation}
+ \tilde{\ket{\phi_1}} \tilde{\bra{\phi_0}} \otimes \ket{1} \bra{0} + \tilde{\ket{\phi_1}} \tilde{\bra{\phi_1}} \otimes \ket{1} \bra{1} .
\label{eq:density}
\vspace{-0.05in}
\end{equation}
For Bob's reduced state to be maximally-mixed across the Alice-Bob partition, the following constraint must be satisfied:
\vspace{-0.13in}
\begin{equation}
\rho_B = \text{Tr}_{123} \left( \rho \right) = \frac{\mathbb{I}}{2}. 
\label{eq:perfect_w_condition}
\vspace{-0.05in}
\end{equation}
The reduced density matrix for Bob's subsystem can be obtained by taking the partial trace over Alice's qubits in Eq.~\ref{eq:density}, which provides:
\vspace{-0.05in}
\begin{equation}
\rho_B = \left( \left| a \right|^2 + \left| b \right|^2 + \left| c \right|^2 \right)  \ket{0} \bra{0} + \left| d \right|^2 \ket{1} \bra{1}.
\label{eq:bobs_reduced_density}
\vspace{-0.05in}
\end{equation}
By defining $p = \left| a \right|^2 + \left| b \right|^2 + \left| c \right|^2$, normalization then requires $1- p = \left| d \right|^2$. We rewrite Eq.~\ref{eq:bobs_reduced_density} as:
\vspace{-0.05in}
\begin{equation}
\rho_B = p \ket{0} \bra{0} + (1 - p) \ket{1} \bra{1}.
\label{eq:bobs_reduced_density_2}
\vspace{-0.05in}
\end{equation}
By setting $p = \frac{1}{2}$, we can satisfy Eq.~\ref{eq:perfect_w_condition}:
\vspace{-0.05in}
\begin{equation}
    \rho_B = \frac{1}{2} \ket{0} \bra{0} + \frac{1}{2} \ket{1} \bra{1}= \frac{\mathbb{I}}{2}.
    \vspace{-0.05in}
\end{equation}
Therefore, we must choose amplitudes such that $p = \left| a \right|^2 + \left| b \right|^2 + \left| c \right|^2 = \frac{1}{2}$ and $1- p = \left| d \right|^2 = \frac{1}{2}$. The perfect $W_4$ resource can thus be defined via the amplitudes $a = b = c = 1 / \sqrt{6}$, $d = 1 / \sqrt{2}$~\cite{pradhan2007teleportationsuperdensecodinggenuine}:
\vspace{-0.05in}
\begin{equation}
\tilde{\ket{W_4}} = \frac{1}{\sqrt{6}} \left( \ket{1000} + \ket{0100} + \ket{0010} + \sqrt{3} \ket{0001} \right)
\label{eq:explicit_perfect_w4_state}
\vspace{-0.05in}
\end{equation}
Since $\tilde{\ket{\phi_0}}$ and $\tilde{\ket{\phi_1}}$ are orthogonal, $\tilde{\bra{\phi_0}} \tilde{\phi_0} \rangle = p$, $\tilde{\bra{\phi_0}} \tilde{\phi_1} \rangle =\tilde{\bra{\phi_1}} \tilde{\phi_0} \rangle = 0$, and $\tilde{\bra{\phi_1}} \tilde{\phi_1} \rangle = 1 - p$. Thus, we can define normalized logical qubits for Alice's subsystem:
\vspace{-0.05in}
\begin{equation}
\ket{0}_L \equiv\ket{\phi_0}_{123} = \frac{\tilde{\ket{\phi_0}}}{\sqrt{p}}, \quad \ket{1}_L \equiv \ket{\phi_1}_{123} = \frac{\tilde{\ket{\phi_1}}}{\sqrt{1 - p}},
\label{eq:logical_qubits}
\vspace{-0.05in}
\end{equation}
with which her four-dimensional logical Hilbert space is $\mathcal{H}_L = \text{span}\{ \ket{0}_A \otimes \ket{\phi_0}, \ket{0}_A \otimes \ket{\phi_1}, \ket{1}_A \otimes \ket{\phi_0}, \ket{1}_A \otimes \ket{\phi_1} \}$. Using these and the constraint that $p = \frac{1}{2}$, it can be seen that the perfect $W_4$ resource is logically equivalent to a Bell state:
\vspace{-0.05in}
\begin{equation}
\tilde{\ket{W_4}} = \ket{\Phi_0^+}_{LB} = \frac{1}{\sqrt{2}} \left( \ket{0}_L \otimes \ket{0}_B + \ket{1}_L \otimes \ket{1}_B \right).
\label{eq:logical_w4_bell_state}
\vspace{-0.05in}
\end{equation}
This form is naturally suited for teleportation, as the two-dimensional logical subspace spanned by $\ket{\phi_0}$ and $\ket{\phi_1}$ can be decoded onto a single logical qubit. These normalized logical basis states are equivalent for both resource states:
\vspace{-0.05in}
\[
    \ket{\phi_0}_{123} = \frac{1}{\sqrt{3}} \left( \ket{100} + \ket{010} + \ket{001} \right), \quad \ket{\phi_1}_{123} = \ket{000}.
    \vspace{-0.05in}
\]
Inspired by the contributions of Agrawal and Pati~\cite{Agrawal_Pati_2006}, we now define a resource-independent orthonormal basis for the four-dimensional logical subspace containing Alice's input qubit and her logical resource qubit:
\vspace{-0.05in}
\begin{equation}
\ket{\eta^{\pm}}_{A123} = \frac{1}{\sqrt2} \left( \ket{0}_A \otimes \ket{\phi_0}_{123} \pm \ket{1}_A \otimes \ket{\phi_1}_{123} \right),
\vspace{-0.05in}
\end{equation}
\begin{equation}
\ket{\xi^{\pm}}_{A123} = \frac{1}{\sqrt2} \left( \ket{0}_A \otimes \ket{\phi_1}_{123} \pm \ket{1}_A \otimes \ket{\phi_0}_{123} \right).
\vspace{-0.05in}
\end{equation}
We define her von Neumann projective measurement as $ \Pi_{m \in \Omega} = \{ \Pi_{00}, \Pi_{01}, \Pi_{10}, \Pi_{11} \}$, where
\vspace{-0.05in}
\[
\Pi_{00} = \ket{\eta^+} \bra{\eta^+}, \quad \Pi_{01} =  \ket{\xi^+} \bra{\xi^+},
\vspace{-0.05in}
\]
\begin{equation}
    \Pi_{10} =  \ket{\eta^-} \bra{\eta^-}, \quad \Pi_{11} =  \ket{\xi^-} \bra{\xi^-},
    \vspace{-0.05in}
\end{equation}
form a complete projective measurement on $\mathcal{H}_{L}$, as $\sum_{m \in \Omega} \Pi_m = \mathbb{I}_{\mathcal{H}_L}$. Expanding the pre-measurement initial composite state in this basis gives the useful expression:
\vspace{-0.05in}
\[
\ket{\psi}_A \otimes \tilde{\ket{W_4}}_{123B} = \frac{1}{2} \left( \ket{\eta^+}_{A123} \otimes \ket{\psi}_B \right.
\vspace{-0.05in}
\]
\[
+ \ket{\xi^+}_{A123} \otimes X\ket{\psi}_B + \ket{\eta^-}_{A123} \otimes Z \ket{\psi}_B
\vspace{-0.05in}
\]
\begin{equation}
\left. + \ket{\xi^-}_{A123} \otimes XZ \ket{\psi}_B \right).
\label{eq:teleportation_identity}
\vspace{-0.05in}
\end{equation}
Alice can thus perform her mid-circuit measurement $\Pi_{m}$ and use classical feed-forward to send the outcome $m = m_z m_x \in \Omega$ she observes to Bob, enabling him to apply the corresponding correction $U_m$ listed in Table~\ref{tab1} and recover the teleported state. The resulting ideal teleportation channel is the identity for the perfect $W_4$ resource, and for the symmetric $W_4$ resource is a dephasing channel.
\begin{table}
\begin{center}
\caption{Measurement Outcomes and Corresponding Pauli Corrections Used in the Branch-Resolved Teleportation Protocol.}
\label{tab1}
\begin{tabular}{cccc}
    \hline
    Alice's Measurement & $m$ & Bob's Correction  \\
    \hline
    $\ket{\eta^+}$ & 00 & $\mathbb{I}$ \\ 
    \hline
    $\ket{\xi^+}$ & 01 & $X$ \\
    \hline
    $\ket{\eta^-}$ & 10 & $Z$ \\
    \hline
    $\ket{\xi^-}$ & 11 & $XZ$ \\
    \hline
\end{tabular}
\end{center}
\vspace{-0.2in}
\end{table}

\subsection{Branch-Resolved Choi Operators}
\label{sec:branch_resolved_choi_operators}

To obtain normalized branch Choi states, we maximally entangle a reference qubit, labeled $R$, with Alice's input qubit $A$ by preparing them in a normalized Bell state:
\vspace{-0.05in}
\begin{equation}
    \ket{\Phi_0^+}_{RA} = \frac{1}{\sqrt{2}} \left( \ket{0}_{R} \otimes \ket{0}_A + \ket{1}_{R} \otimes \ket{1}_A \right) 
\label{eq:bell_state_ra}
\vspace{-0.05in}
\end{equation}
Due to the Choi-Jamio{\l}kowski isomorphism~\cite{Choi_1975, Leung_2003}, the post-teleportation joint operator on the reference qubit and Bob's output qubit is the branch Choi operator $\mathcal{C}_m$ corresponding to each $\tilde{\mathcal{E}}_m$:
\vspace{-0.05in}
\begin{equation}
    \mathcal{C}_m = \left( \mathbb{I}_R \otimes \tilde{\mathcal{E}}_m \right) \left[ \ket{\Phi_0^+} \bra{\Phi_0^+} \right],
\label{eq:branch_choi_operator}
\vspace{-0.05in}
\end{equation}
where $m \in \Omega$ is the true branch induced by the mid-circuit measurement. Given that $\tilde{\mathcal{E}}_m$ is both CP and TNI, it follows that $\mathcal{C}_m$ is positive-semidefinite and subnormalized. In the remainder of this paper, we use the term \emph{branch Choi operator} to refer to the subnormalized Choi operator $\mathcal{C}_m$ associated with a single measurement branch. The aggregate corrected Choi operator for the CPTP teleportation process can be expressed as their sum:
\vspace{-0.05in}
\begin{equation}
    \mathcal{C} = \sum_{m \in \Omega} \mathcal{C}_m
    \vspace{-0.05in}
\end{equation}
For our entangled-input teleportation process, we define the branch probability $p_m$ as the probability of obtaining the true branch $m$ from Alice's mid-circuit measurement. This can be obtained from taking the trace of the branch Choi operator:
\vspace{-0.05in}
\begin{equation}
    p_m = \text{Tr} \left( \mathcal{C}_m \right)
    \vspace{-0.05in}
\end{equation}
The normalized branch Choi states for the post-teleportation reference-output subsystem can be obtained via division of each Choi operator $\mathcal{C}_m$ by the respective branch probability:
\vspace{-0.05in}
\begin{equation}
    \sigma_m = \frac{\mathcal{C}_m}{p_m}
    \vspace{-0.05in}
\end{equation}
We define the operator $M_m = d \left[ \text{Tr}_B \left( \mathcal{C}_m \right) \right]^T$, which is the branch effect operator associated with the CP-TNI branch map in the Choi representation, and the set $\{ M_m \}_{m \in \Omega}$ forms the corresponding Positive Operator-Valued Measure (POVM)~\cite{Davies_Lewis_1970, Leung_2003}.
Here, $d$ is the dimensionality of the input state's Hilbert space. For qubit-based experiments, $d = 2$. These $M_m$ allow us to obtain the conditional probability of obtaining a branch $m$ given an input state $\rho_A$:
\vspace{-0.05in}
\begin{equation}
    p \left( m \mid \rho_A \right) = \text{Tr} \left( M_m \rho_A \right)
    \vspace{-0.05in}
\end{equation}
Although both the symmetric $W_4$ and perfect $W_4$ resource states allow the same logical basis and outcome-dependent corrections, their differing ideal channels cause them to have distinct ideal corrected branch maps and ideal aggregate channels. While we are interested in reconstructing $p_m$, $\sigma_m$, and $\mathcal{C}$ from the true branch Choi operators $\mathcal{C}_m$, the measurement noise obscures our ability to obtain knowledge of the true branch induced by the measurement. It is necessary to instead use a separate label $\ell$ for the observed branch readouts, whose meaning depends on the correction strategy.  

\subsection{Classical Choi Shadows}

Completely characterizing dynamical quantum processes is computationally intensive, as the number of measurements required to do so scales exponentially~\cite{Levy_Luo_Clark_2024, Huang_Kueng_Preskill_2020, volya2024fast} with the number of qubits involved. Our goal experimentally is to estimate the branch-resolved Choi operators $\mathcal{C}_m$ for single-qubit teleportation, and thus the Choi subsystem we aim to reconstruct consists only of the two reference-output qubits $R, B$. While full tomography of two qubits is relatively tractable computationally, even small increases to the number of qubits would quickly render the task impractical for many experimental scenarios. Alternatively, ShadowQPT~\cite{Kunjummen_Tran_Carney_Taylor_2023, Levy_Luo_Clark_2024} has demonstrated that classical-shadow methods can be used to construct stochastic Choi estimators which estimate selected properties of quantum processes using fewer measurements than conventional full process tomography. Thus, we use our two-qubit Choi environment to experimentally validate the accuracy of estimations produced via our non-mitigated Choi shadow estimator (Eq.~\ref{eq:non_mitigated_estimator}) against full process tomography of the branch Choi operators. For multi-qubit teleportation, and experiments with larger reference-output systems, full tomography of these branch Choi operators would quickly become infeasible, in which case our shadow implementation serves as an insightful reference for efficiently estimating appropriately selected, actionable process information without exhaustive process tomography.


We take a branch-resolved approach to performing classical shadow tomography of the post-teleportation branch Choi operators. For each shadow circuit, a randomized two-qubit Pauli basis $b = (b_R, b_B) \in \{ X, Y, Z \}^{\otimes 2}$ for the qubits $R$, $B$ is generated. Each shot of a given shadow circuit then produces: (1) The classical branch readout obtained from the mid-circuit measurement, and (2) A terminal measurement outcome bitstring $s = (s_R, s_B) \in \{0, 1\}^{\otimes 2}$ for qubits $R$, $B$. From these, we construct a two-qubit inverted local Pauli snapshot operator $\tau_{b,s}$ for each shot on qubits $R, B$~\cite{Huang_Kueng_Preskill_2020}:
\vspace{-0.05in}
\begin{equation}
    \tau_{b,s} = \bigotimes_{i \in \{ R, B \}} \left( 3 \, \Pi_{s_i}^{b_i} - \mathbb{I}_i \right)
\label{eq:shadow_snapshot_operator}
\vspace{-0.05in}
\end{equation}
Here, $\Pi_{s_i}^{b_i} = U_{b_i}^{\dagger} \ket{s_i} \bra{s_i} U_{b_i}$ is the rank-1 projector associated with the terminal measurement outcome $s_i$ in the Pauli basis $b_i$, and $U_{b_i}$ is the single-qubit unitary used to perform the measurement in the $b_i$ basis. An important distinction is that for the physical application approach to corrections, the terminal measurement is able to directly measure the teleported output state on $B$. However, for the post-processing approach, no physical corrective gate is applied following the mid-circuit measurement, and the correction is instead applied through tracked branch-dependent operators in post-processing. Thus, the aggregate estimator built from these snapshots is able to estimate the corrected branch processes, regardless of correction strategy.

We define $\ell_i$ as the branch label associated with shot~$i$. The shot-specific snapshot operators are then aggregated into branch-specific shadow estimators $\hat{\mathcal{C}}_{\ell}$ which we use to reconstruct the observed branch Choi operator:
\vspace{-0.05in}
\begin{equation}
    \hat{\mathcal{C}}_{\ell} = \frac{1}{N} \sum_{i = 1}^N \mu_i \delta_{\ell_i, \ell} \,\tau_{b_i,s_i}
\label{eq:non_mitigated_estimator}
\vspace{-0.05in}
\end{equation}
This estimator groups snapshots by branch via the Kronecker delta $\delta_{\ell_i, \ell}$. The $\mu_i$ term is a PROM-specific weight factor we introduce for each snapshot estimator. For our mitigation-free approaches to corrections, we group branches directly by the reported label obtained from mid-circuit measurement, and set $\mu_i$ to 1 for each shot. For PROM mitigation, it is determined by the calibrated quasi-probability coefficients and the corresponding PROM sampling distribution.

\subsection{PROM Mitigation and a Mitigated Branch Estimator}

While the shadow estimator $\hat{\mathcal{C}}_{\ell}$ allows the Choi operator to be approximated, unfortunately, readout noise obscures our ability to directly determine the true branch $m$ obtained from each mid-circuit measurement, regardless of correction strategy~\cite{Koh_Koh_Thompson_2026}. For the unmitigated correction strategies, we only have the reported branch label to work with, and cannot target the ideal branches with our estimator. Alternatively, we can utilize Bit-Flip Averaging (BFA) to symmetrize the effective readout channel~\cite{Smith_Khosla_Self_Kim_2021} through specific device calibration circuits and obtain an error syndrome distribution that, combined with PROM quasi-probability re-weighting, allows us to construct an estimator that uses the calibrated symmetrized readout model to target the ideal branch in its approximations.

To calibrate the effective readout error channel for the qubits $A$ and $3$ which we perform Alice's mid-circuit measurement on, we first define both a true branch label $t = t_z t_x \in \Omega$ and a bit-flip bitstring $u = u_z u_x \in \Omega$, where the subscripts $z$ and $x$ refer to the $Z$ and $X$ components of the correction to the reported branch label, respectively. We use $t$ as a pre-determined label for the true branch we expect from measurement, and $u$ as a known set of bit flips we apply to qubits $A$ and $3$ before measurement with equal probability. If either $u_z$ or $u_x$ is 1, an $X$ gate is applied to qubit $A$ or $3$, respectively. We use a total of 16 circuits to calibrate over all combinations of $(t, u) \in \Omega \times \Omega$. After the mid-circuit measurement in each circuit, we perform modulo-2 addition (equivalently an XOR operation) between the reported branch label $r$ and $u$ to obtain $\tilde{r} = r \oplus u$. We use this to construct the corresponding readout error syndrome $g = \tilde{r} \oplus t$. 
From these calibration circuits, we estimate the 4-element symmetrized error-syndrome distribution $q_g$, which gives the calibrated probability of error syndrome $g$ for the symmetrized mid-circuit measurement on qubits $A$ and $3$:
\vspace{-0.05in}
\begin{equation}
    q_g = \frac{1}{|\Omega|^2} \sum_{t, u \in \Omega} \text{p} \!\left( \left( \tilde{r} \oplus t \right) = g \mid t, u\right)
    \vspace{-0.05in}
\end{equation}
The first element $q_{00}$ is the probability of the readout-error-free syndrome $g = 00$, with corresponding basis vector $e_{00} = (1, 0, 0, 0)^T$. Because we have symmetrized these readout errors, the logical operation of the mid-circuit measurement is maintained~\cite{Koh_Koh_Thompson_2026}. The PROM bitmask is defined as $f \in \Omega$, which is applied to qubit $B$ after the feed-forward correction is made. Our syndrome distribution is then used to construct the $4 \times 4$ symmetrized readout confusion matrix $Q_{g,f} = q_{g \oplus f}$~\cite{Koh_Koh_Thompson_2026}.
which allows us to solve the linear equation $Q\alpha = e_{00}$ and obtain the weight vector $\alpha = (\alpha_{00}, \alpha_{01}, \alpha_{10}, \alpha_{11})^T$ of possibly negative PROM quasi-probability coefficients $\alpha_f$. We use the corresponding $L_1$ norm $\xi = ||\alpha||_1 = \sum_{f \in \Omega} |\alpha_f|$ to obtain the non-negative sampling distribution $p(f)$, from which we sample PROM bitmasks~\cite{Koh_Koh_Thompson_2026}:
\vspace{-0.10in}
\begin{equation}
    p(f) = \frac{|\alpha_f|}{\xi}
    \vspace{-0.06in}
\end{equation}
In our mitigated approach to corrections, we use our experimentally-calibrated PROM coefficients $\alpha_f$ to define the PROM weight for each shot:
\vspace{-0.05in}
\begin{equation}
    \mu_i = \frac{\alpha_{f_i}}{p(f_i)}.
\label{eq:prom_shot_weight}
\vspace{-0.05in}
\end{equation}
After the mid-circuit measurement, we use the reported label $r$ to apply the same correction as in the physical-correction strategy. To complete the BFA Pauli twirl, we apply $X_B^{u_x} Z_B^{u_z}$ to Bob's output qubit, followed by the sampled PROM bitmask $f$. For shot $i$, this produces an effective correction label $c_i = r_{i} \oplus u_i \oplus f_i$, which we use as the branch label $\ell_i$ for grouping our shadow snapshots. Thus, using the calibrated symmetrized readout model, we form our mitigated branch estimator:
\vspace{-0.05in}
\begin{equation}
    \hat{\mathcal{C}}_{\ell}^{\,\text{PROM}} = \frac{1}{N} \sum_{i = 1}^N \frac{\alpha_{f_i}}{p(f_i)} \delta_{\ell_i, \ell}\tau_{b_i, s_i}
\label{eq:mitigated_estimator}
\vspace{-0.05in}
\end{equation}
which approximates mitigated branch Choi operators, indexed by the ideal branch labels.

\subsection{Evaluation Metrics}
\label{sec:evaluation_metrics}

Having established separate shadow estimators $\hat{\mathcal{C}}_{\ell}$ and $\hat{\mathcal{C}}_{\ell}^{\text{PROM}}$ of the corrected branch Choi operators for our unmitigated and mitigated correction strategies, respectively, the common branch label $\ell$ thus has a different meaning depending on the correction strategy. For unmitigated approaches, it is the reported branch label $r$ obtained from Alice's mid-circuit measurement. For the mitigated approach, it represents the corrected label $\ell \equiv c = r \oplus u \oplus f$. Thus, we now define the quantities we reconstruct from these $\hat{\mathcal{C}}_{\ell}$. The estimated branch probabilities are obtained directly from taking the trace:
\vspace{-0.05in}
\begin{equation}
    \hat{p}_{\ell} = \text{Tr} \left( \hat{\mathcal{C}}_{\ell} \right)
    \vspace{-0.05in}
\end{equation}
We denote the ideal normalized corrected branch Choi state as $\omega_{\ell}$. We define the estimated normalized branch Choi state $\hat{\omega}_{\ell}$, and the estimated branch quality $\hat{q}_{\ell}$, respectively, as:
\vspace{-0.05in}
\begin{equation}
    \hat{\omega}_{\ell} = \frac{\hat{\mathcal{C}}_{\ell}}{\hat{p}_{\ell}}, \quad \hat{q}_{\ell} = \frac{\text{Tr} \left( \omega_{\ell} \, \hat{\mathcal{C}}_{\ell} \right) }{\hat{p}_{\ell}}
\label{eq:estimated_branch_quality}
\vspace{-0.05in}
\end{equation}
The branch quality $\hat{q}_{\ell}$ defines the overlap between our experimentally-reconstructed normalized branch Choi state and the corresponding ideal corrected normalized branch Choi state. We also define the branch-resolved feed-forward penalty:
\vspace{-0.13in}
\begin{equation}
    \Delta_{\ell}^{\text{FF}} = \hat{q}_{\ell}^{\text{post}} - \hat{q}_{\ell}^{\text{phys}}
\label{eq:branch_feedforward_error}
\vspace{-0.05in}
\end{equation}
which is the defining metric we use to characterize the observed feed-forward error. It is the difference in branch qualities obtained from the physical application approach to corrections and the post-processing approach, where the latter uses feed-forward only classically in post-processing. Positive values indicate that the physical corrective gate has obscured information related to a given measurement-induced branch.

We define the aggregate feed-forward penalty $\Delta^{\text{FF}}$ as the weighted sum of each branch-specific feed-forward penalty: 
\vspace{-0.02in}
\begin{equation}
    \Delta^{\text{FF}} = \sum_{\ell} \hat{p}_{\ell}^{\text{post}} \Delta_{\ell}^{\text{FF}}
\label{eq:total_feedforward_penalty}
\vspace{-0.05in}
\end{equation}
This gives the cumulative branch-quality penalty associated with physical feed-forward operations, weighted by the experimentally-estimated post-processing branch probabilities. Finally, we evaluate the extent of mitigation achieved from PROM within each branch by comparing the PROM branch qualities with those of physical application:
\vspace{-0.03in}
\begin{equation}
    G_{q, \ell} = \hat{q}_{\ell}^{\text{prom}} - \hat{q}_{\ell}^{\text{phys}}
\label{eq:prom_branch_mitigation}
\vspace{-0.04in}
\end{equation}
Positive values indicate that PROM improves the reconstructed branch qualities relative to physical correction. In our aggregate metrics plot, we describe the average PROM mitigation across all branch qualities using the following unweighted mean over all four branch labels:
\vspace{-0.02in}
\begin{equation}
    \bar{G}_{q} = \frac{1}{4} \sum_{\ell \in \Omega} G_{q, \ell}
\label{eq:mean_branch_mitigation}
\vspace{-0.03in}
\end{equation}
We compare the accuracy of our Choi-shadow estimates to full Choi-state tomography using two families of scalar observables. First, we define the estimated branch effect operator:
\vspace{-0.12in}
\begin{equation}
    \hat{M}_{\ell} = d \left[ \text{Tr}_B \left( \hat{\mathcal{C}}_{\ell} \right) \right]^T
    \vspace{-0.05in}
\end{equation}
where $d = 2$ is the Hilbert space dimension for the input qubit. For each branch $\ell \in \Omega$, our primary set of observables is:
\vspace{-0.03in}
\begin{equation}
    \mathcal{O}_{\ell}^{\text{Primary}} = \{ \hat{p}_{\ell}, \hat{M}_{\ell, X}, \hat{M}_{\ell, Y}, \hat{M}_{\ell, Z}, \hat{q}_{\ell} \}
\label{eq:primary_observables}
\vspace{-0.03in}
\end{equation}
where $\hat{M}_{\ell, X}$, $\hat{M}_{\ell, Y}$, and $\hat{M}_{\ell, Z}$ are the non-trivial Pauli coefficients of the branch effect operator $\hat{M}_{\ell}$, given by $\hat{M}_{\ell, P} = \text{Tr} ( \!P \hat{M}_{\ell} )$. The second set of observables we use is a complete branch Choi observable family, which we define as:
\vspace{-0.03in}
\begin{equation}
\mathcal{O}^{\text{Full}} = \left\{ \text{Tr} \left[ (P_R \otimes P_B) \hat{\mathcal{C}}_{\ell} \right] :  \ell \in \Omega, \, P_R, P_B \in \mathbb{P} \right\}.
\label{eq:full_observables}
\vspace{-0.03in}
\end{equation}
For both observable families, we compare the shadow estimate with the estimate from full tomography using:
\vspace{-0.03in}
\begin{equation}
\footnotesize
\mathrm{RMSE} = \sqrt{\frac{1}{|\mathcal{K}|} \sum_{k \in \mathcal{K}} \left( \mathcal{O}^{\mathrm{shadow}}_k - \mathcal{O}^{\mathrm{tomography}}_k \right)^2},
\label{eq:rmse}
\vspace{-0.05in}
\end{equation}
where $\mathcal{K}$ is the set of shared scalar observable labels.

\section{Experiments}
\label{sec:experimental_methods}

\subsection{IBM Superconducting Hardware}

All experiments were performed on IBM's 156-qubit Heron r2 superconducting quantum processing unit (QPU) IBM Fez. Experiments were designed in Qiskit software, and submitted to the backend as 10-minute batches via Qiskit runtime service. The fixed transpilation seed 98364 was used for reproducibility. All circuits were transpiled with optimization level 1.

\subsection{Logical and Physical Qubit Layouts}

Following our convention in Section~\ref{sec:Background}, we label the logical qubits in our circuits as $[R, A, 1, 2, 3, B]$, denoting $R$ as the reference qubit, $A$ as the input qubit, $1$, $2$, $3$ as Alice's resource qubits, and $B$ as Bob's qubit. We performed mid-circuit measurements on logical qubits $A$ and $3$ in our implemented circuits, and terminal measurements on qubits $R$ and $B$. The mappings of logical to physical qubits for our layouts are:
\vspace{-0.05in}
\[
\text{Layout 1: } [R, A, 1, 2, 3, B] \rightarrow [7, 6, 3, 4, 5, 16]
\vspace{-0.05in}
\]
\[
\text{Layout 2: } [R, A, 1, 2, 3, B] \rightarrow [22, 23, 2, 3, 16, 1]
\vspace{-0.05in}
\]
The mid-circuit measurement pairs were $(A, 3) = (6, 5)$ for layout 1, and $(A, 3) = (23, 16)$ for layout 2. These were chosen specifically so that the physical qubits mapped to $A$ and $3$ would have high readout assignment error (RAE) for the first layout, and low RAE in the second. The relevant noise characteristics for each qubit were obtained from the official IBM Fez calibration data available at the time we performed our experiments, and are listed in Table~\ref{tab2}.

\begin{table}[htbp]
\begin{center}
\vspace{-0.1in}
\caption{IBM Fez Noise Calibration Data}
\vspace{-0.05in}
\label{tab2}
\begin{tabular}{lcccc}
    \hline
    Physical Qubit  & 5 & 6 & 16 & 23 \\
    \hline
    Readout Assignment Error  & 0.05139 & 0.01831 & 0.00610 & 0.00671 \\ 
    T1 Relaxation Time (us) & 164.684 & 265.535 & 248.926 & 152.267 \\
    T2 Coherence Time (us) & 136.006 & 216.041 & 165.657 & 157.805 \\
    \hline
\end{tabular}
\end{center}
\vspace{-0.15in}
\end{table}

\subsection{Controlled Resource State Preparation}

\begin{figure}[htbp]
\vspace{-0.1in}
  \centering
  \includegraphics[width=1.0\linewidth]{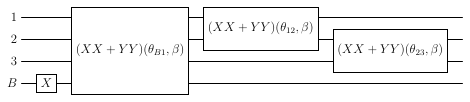}
  \vspace{-0.2in}
  \caption{State preparation quantum circuit. Prepares either entangled $W_4$ resource state with matched circuit depth and gate count. $\beta = \pi / 2$, and $\theta_{B1} = \pi / 2$ prepares the perfect $W_4$ resource, while $\theta_{B1} = 2\pi / 3$ prepares the symmetric $W_4$ resource. $\theta_{12}$ and $\theta_{23}$ are the same for both resources, as shown in Table~\ref{tab3}.}
  \label{fig:resource_prep_circuit}
  \vspace{-0.2in}
\end{figure}

We first prepare the explicit physical $W_4$ resource states described in Section~\ref{sec:Background} on qubits $ 1, 2, 3, B$ with equal circuit depth and gate counts. By applying an $X$ gate to qubit $B$, and then distributing the resulting amplitude across qubits $1, 2, 3$ via the parameterized 2-qubit gates shown in Figure~\ref{fig:resource_prep_circuit}, we obtain the explicit form of either the symmetric state given in Eq.~\ref{eq:explicit_symmetric_w4} or the perfect $W_4$ state given in Eq.~\ref{eq:explicit_perfect_w4_state}, depending on the choice of parameter $\theta_{B1}$. The exact parameters are provided in Table~\ref{tab3}.

In our teleportation protocol, we perform Alice's measurement as a Bell measurement on qubits $A$ and $3$. Thus, after preparing the explicit forms of the resource states on qubits $1, 2, 3, B$ via the quantum circuit in Fig.~\ref{fig:resource_prep_circuit}, we decode the resource state qubits into their logical basis (Eq.~\ref{eq:logical_qubits}) so that the two-dimensional logical subspace on qubits $1, 2, 3$ is mapped to a single logical qubit on $3$. This is accomplished through the decoding circuit shown below in Fig~\ref{fig:logical_decode}, and allows Alice's von Neumann measurement $\Pi_m$ to be implemented experimentally as a normal Bell measurement on qubits $A$ and $3$.

\begin{table}[htp]
\vspace{-0.1in}
\begin{center}
\caption{Parameters for state preparation XX + YY gates in Fig.~\ref{fig:resource_prep_circuit}.}
\label{tab3}
\begin{tabular}{lccc}
    \hline
    Entanglement Resource & $\theta_{B1}$ & $\theta_{12}$ & $\theta_{23}$ \\
    \hline
    Perfect $W_4$ & $\pi / 2$ & $2\arccos(1 / \sqrt{3})$ & $\pi / 2$ \\
    Symmetric $W_4$ & $2\pi / 3$ & $2\arccos(1 / \sqrt{3})$ & $\pi / 2$ \\
    \hline
\end{tabular}
\end{center}
\vspace{-0.1in}
\end{table}

The state preparation circuit in Fig.~\ref{fig:resource_prep_circuit} is transpiled together with the logical decoding circuit in Fig.~\ref{fig:logical_decode}, with a combined circuit depth of 36 and size of 74, totaling 36 $RZ$ gates, 28 $SX$ gates, and 10 $CZ$ gates. These depths and counts are the same for both resource states, which allows them to be used as a controlled comparison for analyzing how the differing degrees of entanglement affect the observed error behavior. Once the entangled resource is prepared, we prepare the entangled Bell state in Eq.~\ref{eq:bell_state_ra} on qubits $R$ and $A$. We prepare the logical resource states first in order to minimize the extent of idle decoherence experienced by the teleportation input state preceding Alice's measurement.

\begin{figure}[htbp]
  \centering
      \vspace{-0.1in}
  \includegraphics[width=1.0\linewidth]{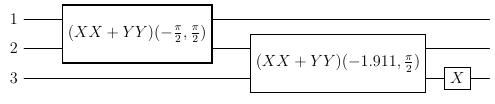}
  \vspace{-0.3in}
  \caption{Decoding quantum circuit. Decodes Alice's resource qubits into a single logical qubit on qubit $3$.}
  \label{fig:logical_decode}
    \vspace{-0.1in}
\end{figure}

\subsection{Teleportation Experiments}

For each layout, we perform three sets of experiments corresponding to each of the three correction strategies: (i) physical correction, (ii) post-processing, and (iii) PROM mitigation. Regardless of the strategy, each shadow circuit begins with resource preparation, followed by preparation of a Bell state on qubits $R$ and $A$. A Bell-basis change is then performed on qubits $A$ and $3$, followed by a mid-circuit measurement which produces a reported branch label $r = r_z r_x \in \Omega$. The first bit $r_z$ corresponds to the $Z$-component of the experimental correction $U_r = X_B^{r_x}Z_B^{r_z}$, and the second bit $r_x$ corresponds to the $X$-component. For all three strategies, one randomized 2-qubit terminal Pauli measurement basis $b = (b_R, b_B)$ is sampled for each shadow circuit, with each circuit executed for 64 shots. The result of each terminal measurement is labeled $s = (s_R, s_B)$. The strategy-dependent operations for each shadow circuit are explained below:

\paragraph{Physical Correction}

Immediately following Alice's (Bell) measurement $\Pi_r$, the outcome $r = r_z r_x$ is recorded classically and used to apply the corresponding correction $U_{r}$ physically to qubit $B$. Afterward, a set of basis-change operations are performed if necessary for measuring $R$ and $B$ in bases $b_R$ and $b_B$, respectively. The terminal measurement on qubits $R, B$ is performed, and the outcome $s$ is recorded. The local snapshot in Eq.~\ref{eq:shadow_snapshot_operator} is computed, and the branch label is calculated as $\ell_i = r_i$. Each shot thus contributes $\frac{1}{N}\delta_{\ell_i, \ell} \tau_{b_i, s_i}$ to the non-mitigated estimator (Eq.~\ref{eq:non_mitigated_estimator}), where $N$ is the total number of shots and $\mu_i = 1$.

\paragraph{Post-Processing}

After Alice's mid-circuit measurement, the outcome $r$ is recorded, however the corresponding correction $U_{r}$ is not applied physically, and instead is represented as a tracked Pauli frame $F_r = X_B^{r_x} Z_B^{r_z}$ to be used in post-processing. A pre-measurement basis change is performed according to the randomly-chosen bases $b_R, b_B$, and the terminal measurement is performed. Afterwards, Bob's outcome $s_B$ is flipped depending on (i) the tracked Pauli frame $F_r$, and (ii) Bob's terminal measurement basis $b_B$:
\begin{enumerate}
    \item If $b_B = X$, $s_B$ is flipped via classical relabeling in post-processing, only when the tracked Z-component bit is 1.
    \item If $b_B = Z$, $s_B$ is flipped only when the tracked X-component bit is 1.
    \item If $b_B = Y$, $s_B$ is flipped only when $r_x \oplus r_z = 1$.
\end{enumerate}
The shot-wise contribution of the snapshot to the non-mitigated estimator is calculated and performed the same as for physical applications.

\paragraph{PROM Mitigation}

For each PROM shadow circuit, in addition to the random terminal measurement basis $b = b_{R}b_{B}$, a bit-flip string $u = u_{z} u_{x}$ is assigned from a uniformly shuffled list over $\Omega$, and a PROM bitmask $f = f_{z} f_{x}$ is sampled from the experimentally calibrated PROM sampling distribution $p(f)$. Before Alice's mid-circuit measurement is performed, the measurement qubits $A$ and $3$ are Pauli-twirled via application of $X_A^{u_{z}} X_3^{u_{x}}$ in order to symmetrize the readout error channel. The mid-circuit measurement is then performed, and $r$ is recorded and used to physically apply the correction $U_{r}$. BFA is then completed via the application of $X_B^{u_{x}}Z_B^{u_{z}}$ to Bob's output qubit $B$ before the terminal measurement. The PROM bitmask $f$ is applied to Bob's qubit $B$, with the terminal measurement performed afterward. For each shot $i$, a corrected branch label $c_i = r_{i} \oplus u_i \oplus f_i$ is computed and used as the effective branch label $\ell_i$. The snapshot $\tau_{b_i, s_i}$ is calculated as in the other strategies using Eq.~\ref{eq:shadow_snapshot_operator}; however, the weight is computed according to Eq.~\ref{eq:prom_shot_weight}. Thus, each shot contributes $\frac{1}{N} \frac{\alpha_{f_i}}{p(f_i)} \delta_{\ell_i, \ell} \tau_{b_i, s_i}$ to the mitigated estimator (Eq.~\ref{eq:mitigated_estimator}).

\subsection{Calibration and Drift Tracking}

To account for hardware drift and obtain a meaningful PROM calibration of the error syndrome present in our physical qubit layouts, we partition the experiments into separate batches. For each layout, two batches are performed, with each batch consisting of an initial and a final PROM calibration of the error syndromes of the qubits used for Alice's mid-circuit measurement. Each calibration uses the complete set of 16 BFA circuits needed to calibrate each combination $(t, u) \in \Omega \times \Omega$ described in Section~\ref{sec:Background}, and 2048 shots are executed for each calibration circuit. From these circuits, we obtain the symmetrized readout error syndrome $q_g$ for the qubits $A, 3$, and use it to calculate the PROM quasi-probability coefficients $\alpha_f$ and the experimental bitmask sampling distribution $p(f)$. The $\alpha_f$ used at runtime are determined from the initial calibration, while those used in post-processing are derived from the average of the initial and final calibrations. The same PROM calibration is used for both resources. 

\subsection{Shot Budgets}

A total of 1,146,880 shots were performed in our main teleportation experiments,  which were split into a total of four batches, with two for each physical layout. Within each batch, calibrations were performed both before and immediately after the teleportation experiments. The 16 BFA calibration circuits needed to fully characterize the readout error syndrome were performed for each batch calibration, with 2048 shots per calibration circuit. Thus, 262,144 total shots were used for PROM calibration. Within each physical layout, 147,456 shots were performed for each correction strategy, with 73,728 shots for each resource, as 1152 shadow circuits were performed 64 times each. Thus 442,368 shadow circuit shots were performed per layout, totaling 884,736 shots for the main teleportation experiments. The separate tomography validation experiments on layout 1, performed for both resources and the two unmitigated strategies, contributed an additional $589,824$ shots ($294,912$ for tomography, and $294,912$ for shadow circuits).

\subsection{Tomography Validation of Branch Choi Operators}

To validate the quality of our shadow estimates, we perform full tomography of the branch Choi operators for the physical-application and post-processing correction strategies. In our tomography baseline, we use all nine two-qubit Pauli measurement bases $\{X,Y,Z\}^{\otimes 2}$ on qubits $R$ and $B$, with 8192 shots per basis. For each resource and correction strategy, this gives a total of $9 \times 8192 = 73,728$ tomography shots. Counts are grouped by the reported branch label $\ell$; for the post-processing strategy, Bob's terminal measurement outcome is additionally relabeled according to the tracked Pauli frame.

Rather than using Qiskit-native tomography, we reconstruct each normalized branch state on $R,B$ using a fixed-dilution maximum likelihood state tomography approach inspired by prior diluted maximum likelihood tomography~\cite{Rehacek_Hradil_Knill_Lvovsky_2007}. Given measurement projectors $\Pi_j$, observed counts $n_j$, and total counts $N = \sum_j n_j$, we initialize a full-rank density matrix $\rho_0$. At iteration $k$, we compute outcome probabilities $p_j^{(k)} = \text{Tr} ( \Pi_j \rho_k)$ and form the state-dependent operator $\mathcal{R}(\rho_k)$:
\vspace{-0.05in}
\begin{equation} 
	\mathcal{R}(\rho_k) = \frac{1}{N} \sum_j \frac{n_j}{p_j^{(k)}} \Pi_j 
    \vspace{-0.05in}
\end{equation} 
We update $\rho_k$ as follows:
\vspace{-0.05in}
\begin{equation} 
	\rho_{k + 1} = \frac{G_{k} \rho_{k} G_{k}}{\text{Tr}\left( G_{k} \rho_{k} G_{k}\right)}, \quad G_{k} = \frac{\mathbb{I} + \epsilon \mathcal{R}(\rho_k)}{1 + \epsilon},
    \vspace{-0.05in}
\end{equation} 
using fixed $\epsilon = 1$. Each reconstructed normalized branch density matrix is then multiplied by its empirical branch probability to obtain a branch Choi operator estimate.

We compare these tomography reconstructions with Choi-shadow estimates from eight shadow-circuit budgets: 9, 18, 36, 72, 144, 288, 576, and 1152 executed shadow circuits from the full randomized shadow dataset. Each shadow circuit uses one sampled terminal Pauli basis $b = b_R b_B \in \{X,Y,Z\}^{\otimes 2}$ and is executed with 64 shots, so the total shadow-shot budgets range from 576 to 73,728. For each budget, we compute the RMSE (Eq.~\ref{eq:rmse}) between the shadow and tomography estimates over the primary observable family in Eq.~\ref{eq:primary_observables} and the full branch Choi observable family in Eq.~\ref{eq:full_observables}. We exclude PROM from this comparison because its mitigated estimator targets ideal branch labels through BFA/PROM quasi-probability weighting, so direct reported-label tomography is not a comparable baseline.

The tomography validation was performed only on layout~1. For each intermediate shadow budget, the RMSE is recomputed over 16 random subsets of executed shadow circuits, and the plotted error bars show the resulting standard deviation. At the full 1152 shadow circuit budget, all shadow circuits are used, so the standard deviation is zero.

\subsection{Statistical Uncertainty Estimation} 

To estimate shot noise in our sampled shadow circuits, we perform bootstrap resampling from the empirical outcome distribution of each circuit histogram~\cite{Efron_1979, Efron_Tibshirani_1986}. For each circuit, the sum of the observed counts $\{n_x\}$ is given by $N = \sum_x n_x$, and the empirical probabilities for each outcome are $\hat{p}_x = n_x / N$. For a given shadow circuit with $N$ total shots, classical outcome categories indexed by $x = 1, \ldots, K$, and empirical probabilities $\hat{p}_x$, we use NumPy to generate the $j^{\text{th}}$ bootstrap replicate count vector $\tilde{\textbf{n}}^{(j)} = \left( \tilde{n}_1^{(j)} , \ldots , \tilde{n}_K^{(j)} \right)$ as:
\begin{equation}
    \tilde{\mathbf{n}}^{(j)} \mid \mathbf{n} \sim \text{Multinomial}_{K} \left(N, \hat{\mathbf{p}} \right),
\end{equation}
where $\textbf{n} = \left( n_1, \ldots , n_K \right)$ and $\hat{\mathbf{p}}=(\hat{p}_1,\ldots,\hat{p}_K)$. This is done independently for each circuit histogram in our shadow dataset. For every bootstrap resample, we recalculate the same branch-resolved estimators (Eq.~\ref{eq:non_mitigated_estimator}, Eq.~\ref{eq:mitigated_estimator}) and the derived metrics we plot. The PROM coefficients are held fixed during resampling at their analysis-time calibrated values for each block. In our implementation, we use $B = 200$ bootstrap replicates for each shadow dataset, and a 95\% confidence level $\gamma = 0.95$ which is used to define upper and lower quantiles via $\alpha = (1 - \gamma) / 2$. In our plots, central values are the point estimates from the observed counts, while the error bars indicate the empirical $\alpha$ and $(1 - \alpha)$ quantiles of the bootstrap distribution, corresponding to $95\%$ percentile bootstrap confidence intervals.

\section{Experimental Results}
\label{sec:results}

Since the reference-output subsystem for our teleportation protocol consists of only two qubits ($R, B$), full tomography of the branch Choi operators is computationally tractable and provides a conventional baseline for validating our non-mitigated Choi shadow estimator. Our tomography dataset was acquired from experiments on layout 1 for the physical application and post-processing correction strategies. 
\begin{figure}[H]
  \centering
  \vspace{-0.1in}
  \includegraphics[width=1.0\linewidth]{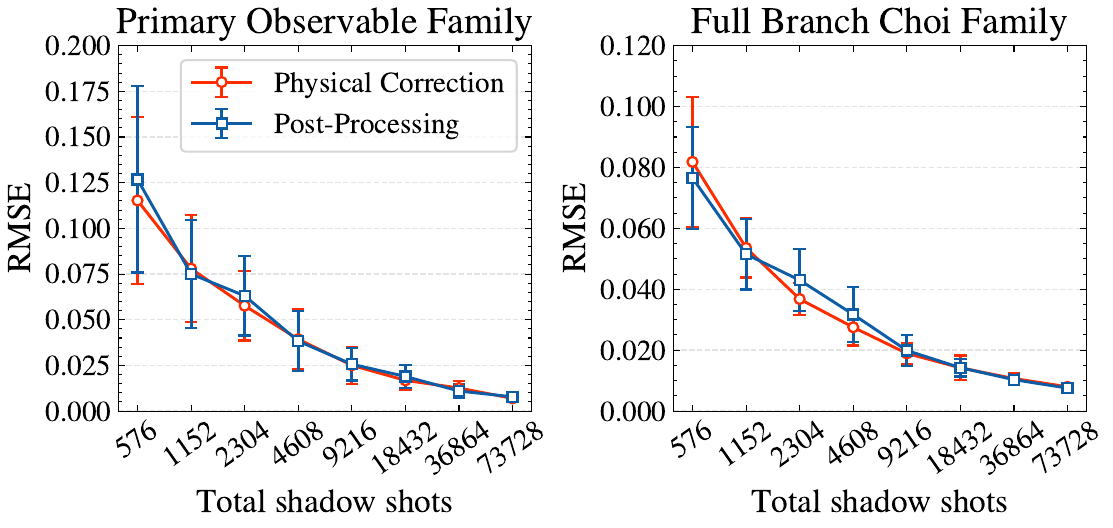}
  \vspace{-0.3in}
  \caption{Shadow estimator validation against full tomography of branch Choi operators on layout 1 using the Perfect $W_4$ resource. Curves show the RMSE between Choi-shadow estimates and the tomography baseline for the physical correction and post-processing strategies as a function of total shadow shots. Left panel: primary observable family $\mathcal{O}_{\ell}^{\text{Primary}}$. Right panel: full branch Choi operator family $\mathcal{O}^{\text{Full}}$. Error bars indicate the standard deviation over 16 random subsets of executed shadow circuits at fixed shot budgets.}
  \label{fig:tomography_validation}
\end{figure}
At the full $73,728$ shot budget shown in Fig.~\ref{fig:tomography_validation}, the perfect $W_4$ RMSE values for the primary observable family are 0.00694 (physical application) and 0.00779 (post-processing). For the full branch Choi observable family, the RMSE values for physical application and post-processing are 0.00797 and 0.00754, respectively. The same experiment was performed with the symmetric $W_4$ state on layout 1, which we do not plot due to page constraints. For the primary observable family, the full-budget RMSE values were 0.01089 (physical application) and 0.01499 (post-processing), and for the full branch Choi observable family, were 0.00838 (physical application) and 0.01108 (post-processing). These results confirm that our non-mitigated Choi-shadow estimator closely matches the full tomography baseline for our two-qubit Choi experiments.

\subsection{Branch-Resolved Feed-Forward Characterization}

\begin{figure}[htbp]
  \centering
  \includegraphics[width=1.0\linewidth]{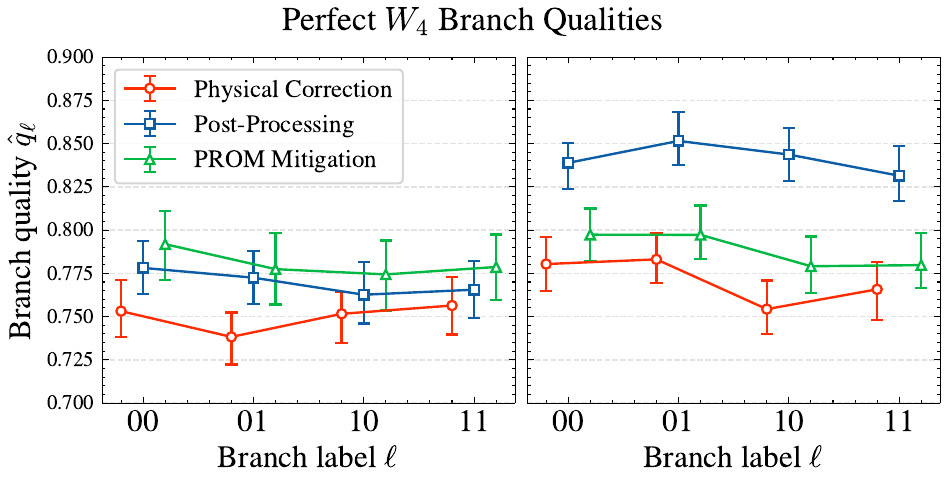}
  \vspace{-0.3in}
  \caption{Estimated branch qualities $\hat{q}_{\ell}$ obtained experimentally on IBM Fez from each of the three correction strategies with the perfect $W_4$ resource. Left: results on layout 1. Right: results on layout 2. Error bars represent 95\% bootstrap confidence intervals. Relative ordering between post-processing and PROM mitigation effectiveness is reversed between the two layouts.}
  \label{fig:branch_quality_perfect_w4}
  \vspace{-0.1in}
\end{figure}

Having established that our Choi-shadow estimator provides accurate reconstructions of the branch Choi operators, we now present our results from using our branch-resolved estimators to study layout-dependent feed-forward error and PROM mitigation in our main controlled-layout teleportation experiments. The same experimental conditions were used for each layout: the number of shadow circuits, shots per shadow circuit, and calibration methods are identical for experiments in both layouts. Branch quality estimates (Eq.~\ref{eq:estimated_branch_quality}) obtained using the perfect $W_4$ resource state are shown in Fig.~\ref{fig:branch_quality_perfect_w4}. In layout 1, PROM mitigation achieves the highest branch qualities for all branches, and physical correction achieves the lowest. Post-processing achieves branch qualities that are worse than PROM but better than physical correction for every branch. In layout 2, however, it is post-processing that achieves the highest branch qualities, exceeding PROM for all branches. Physical correction achieves the lowest branch qualities.

\begin{figure}[hbtp]
  \centering
  \includegraphics[width=1.0\linewidth]{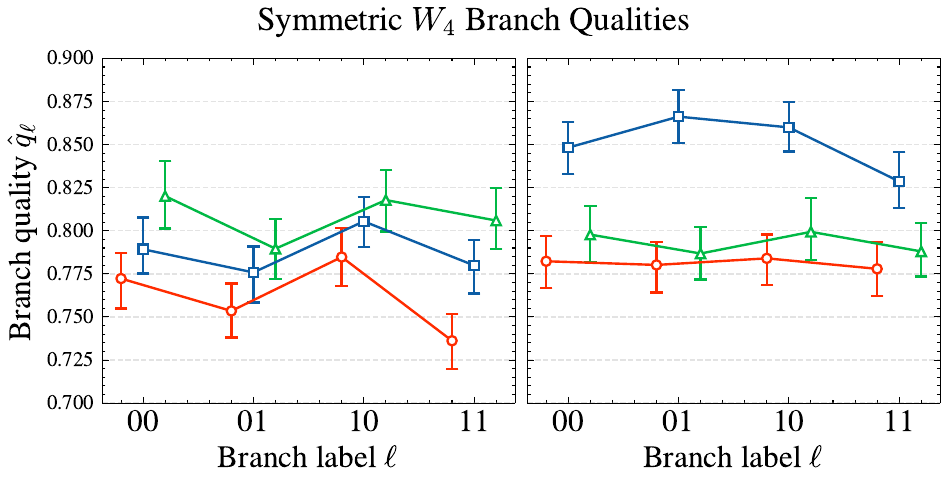}
  \vspace{-0.3in}
  \caption{Estimated branch qualities $\hat{q}_{\ell}$ for all four branch labels $\ell$ obtained using the symmetric $W_4$ resource. Left panel: results from physical qubit layout 1. Right panel: results from physical qubit layout 2. Error bars show the 95\% bootstrap confidence intervals. The layout-dependent reversal between PROM mitigation and post-processing is maintained across both resources.}
  \label{fig:branch_quality_symmetric_w4}
\end{figure}

Branch quality estimates obtained using the symmetric $W_4$ resource state are shown in Fig.~\ref{fig:branch_quality_symmetric_w4}. In layout 1, the same pattern as  Fig.~\ref{fig:branch_quality_perfect_w4} is observed: PROM leads in branch qualities, followed by post-processing, followed by physical correction. In layout 2, similar to Fig.~\ref{fig:branch_quality_perfect_w4}, the relative effectiveness of PROM compared to post-processing is swapped compared to layout 1. Physical correction has the lowest qualities in every branch, just as with the perfect $W_4$ resource. Even though the exact branch qualities differ between the two resource states, the observed layout-determined reversal in effectiveness between PROM mitigation and post-processing is preserved.

\begin{figure}[htbp]
  \centering
  \includegraphics[width=1\linewidth]{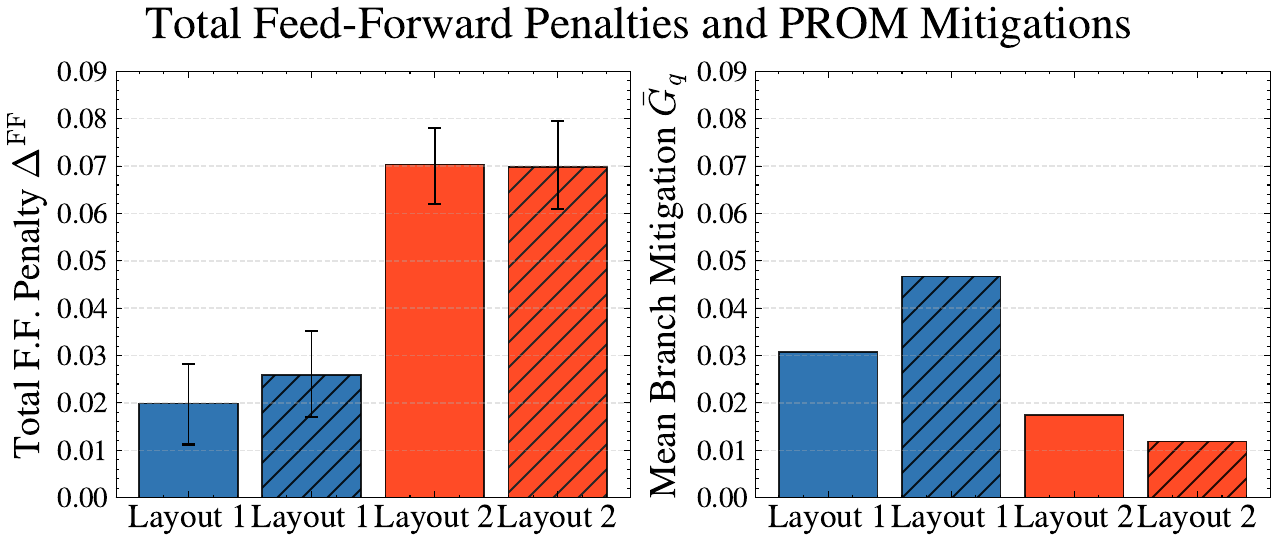}
  \vspace{-0.25in}
  \caption{Total feed-forward (F.F.) penalties $\Delta^{\text{FF}}$ and mean PROM mitigation of branch qualities $\bar{G}_q$ for each resource state and physical qubit layout. Hatched bars refer to results from symmetric $W_4$ state experiments, while unhatched bars refer to perfect $W_4$ results. Error bars in the left panel show the $95\%$ bootstrap confidence intervals. Layout 1 shows lower total feed-forward penalty and noticeably improved PROM mitigation, while layout 2 shows higher total feed-forward penalty and reduced PROM mitigation.}
  \label{fig:feedforward_penalty_prom_gains}
  \vspace{-0.15in}
\end{figure}

Fig.~\ref{fig:feedforward_penalty_prom_gains} summarizes our observed aggregated channel metrics. The left panel shows the total feed-forward penalty $\Delta^{\text{FF}}$ (Eq.~\ref{eq:total_feedforward_penalty}), and the right panel shows the mean PROM mitigation of branch qualities (Eq.~\ref{eq:mean_branch_mitigation}), averaged over all four branches.

For layout 1, the total feed-forward penalties are relatively modest at 0.0199 for the perfect $W_4$ resource, and 0.0259 for the symmetric $W_4$ resource. For layout 2, the total feed-forward penalties become much larger, at 0.0703 for the perfect $W_4$ resource, and 0.0698 for the symmetric $W_4$ resource. The opposite trend is reflected in the mean PROM gain in branch quality $\bar{G}_q$. For layout 1, PROM was able to recover branch qualities much better for each resource state: the perfect $W_4$ resource resulted in a mean PROM gain of 0.0307, while the symmetric $W_4$ state yielded 0.0467. Conversely, layout 2 had reduced PROM mitigation, with mean PROM gains of only 0.0175 with the perfect $W_4$ resource, and 0.0119 for the symmetric $W_4$ resource.

These layout-dependent trends are consistent with our experimentally-obtained PROM calibration data. In our initial and final calibration batches for each layout, the no-error syndrome probability $q_{00}$ is  lower in layout 1 than layout 2. For layout 1, the average calibrated readout error syndrome distribution for the two calibration blocks was $(q_{00}, q_{01}, q_{10}, q_{11}) = (0.9308, \, 0.0509, \, 0.0174, \, 0.0009)$, whereas for layout 2 it was $(q_{00}, q_{01}, q_{10}, q_{11}) = (0.9885, \, 0.0054, \, 0.0061, \, 0.0000)$. Thus, higher levels of PROM mitigation were obtained from the layout with the noisier  readout channel for the mid-circuit measurement. Conversely, the layout with less readout noise produced a much larger separation between post-processing and physical correction branch qualities, providing a clear view of the overhead associated with physical feed-forward.

\section{Conclusion}
\label{sec:conclusion}

In this work, we introduced a framework for characterizing feed-forward error in dynamic teleportation via the reconstruction of branch-resolved Choi operators. We also validated the physical-application and post-processing Choi-shadow reconstructions against full tomography of the branch Choi operators on layout 1, providing a conventional tomographic baseline for the estimator used in the main experiments. Using two physical qubit layouts on IBM Fez, three correction strategies, and two entangled $W_4$ resource states, we found that the level of readout error present in the qubits used for the branch-inducing mid-circuit measurement strongly influences both the cumulative extent of feed-forward error and the relative advantage attained from PROM compared to post-processing. In the layout which has the noisier readout channel, PROM produced the highest branch qualities, and the total feed-forward penalty remained relatively low, while for the layout with the superior readout channel, post-processing produced the highest branch qualities, PROM showed reduced levels of mitigation, and the total feed-forward penalty increased over 2.5 times in magnitude. This pattern persisted across both resource states, demonstrating that our branch-resolved characterization is able to reveal error structure that would be obscured by state-of-the-art outcome-averaged analyses.

\balance

\bibliographystyle{IEEEtran}
\bibliography{references}

\end{document}